\documentclass[sigconf]{acmart}

\usepackage{booktabs} 
\usepackage[colorinlistoftodos,prependcaption,textsize=tiny]{todonotes}
\usepackage{mathtools}
\usepackage{blindtext}
\usepackage{caption}
\usepackage{bbm}
\usepackage{flushend}

\setcopyright{iw3c2w3}

\copyrightyear{2018}
\acmYear{2018} 

\acmConference[WWW '18 Companion]{The 2018 Web Conference Companion}{April 23--27, 2018}{Lyon, France}
\acmBooktitle{WWW '18 Companion: The 2018 Web Conference Companion, April 23--27, 2018, Lyon, France}
\acmPrice{}
\acmDOI{10.1145/3184558.3188724}
\acmISBN{978-1-4503-5640-4/18/04}

\begin{document}

\title{Selection Bias in News Coverage: Learning it, Fighting it}

\author{Dylan Bourgeois}
\authornote{Both authors contributed equally to the paper}
\affiliation{%
  \institution{EPFL}
}
\email{dylan.bourgeois@epfl.ch}

\author{J{\'e}r{\'e}mie Rappaz}
\authornotemark[1]
\affiliation{%
  \institution{EPFL}
}
\email{jeremie.rappaz@epfl.ch}

\author{Karl Aberer}
\affiliation{%
  \institution{EPFL}
}
\email{karl.aberer@epfl.ch}

\newcommand{\xhdr}[1]{\vspace{2mm}\noindent{{\bf #1}}}

\begin{abstract}

News entities must select and filter the coverage they broadcast through their respective channels since the set of world events is too large to be treated exhaustively. The subjective nature of this filtering induces biases due to, among other things, resource constraints, editorial guidelines, ideological affinities, or even the fragmented nature of the information at a journalist's disposal. The magnitude and direction of these biases are, however, widely unknown. The absence of ground truth, the sheer size of the event space, or the lack of an exhaustive set of absolute features to measure make it difficult to observe the bias directly, to characterize the leaning's nature and to factor it out to ensure a neutral coverage of the news.

In this work, we introduce a methodology to capture the latent structure of media's decision process on a large scale. Our contribution is multi-fold. First, we show media coverage to be predictable using personalization techniques, and evaluate our approach on a large set of events collected from the GDELT database. We then show that a personalized and parametrized approach not only exhibits higher accuracy in coverage prediction, but also provides an interpretable representation of the selection bias. Last, we propose a method able to select a set of sources by leveraging the latent representation. These selected sources provide a more diverse and egalitarian coverage, all while retaining the most actively covered events. 

\end{abstract}



\maketitle

\section{Introduction}
\label{sec:intro}
World events are reported through an ever increasing number of information channels. These events happen on a variety of different scales, from global to highly local, and all across the planet. To get a grasp on the world's state, even avid readers must pre-process the event space, with such sampling inherently exposing them to a distorted perspective. This processing is a conscious \textit{selection} which not only applies to the final consumer (the reader) but also to the provider: news sources.

News organizations are designed to be the initial filter of the event stream, pruning, condensing and categorizing it into manageable chunks of information. Unfortunately, it is difficult to guarantee the neutrality of this selection: the process is performed by the editorial team based on an arbitrary number of factors. Some of them are obvious, such as geographic considerations, editorial guidelines, thematic regards or even logistic capabilities. Others are not visible at a glance: ideological leanings or higher order structures such as broadcast syndications or corporate structures. Either one of these can compromise the representativity of the news sample presented: this is generally referred to as \textit{gatekeeping} or \textit{selection bias}.

Any attempt to measure the influence of these factors on news coverage in absolute terms is ill-fated: the factor space could never claim to be exhaustive, and a subset would be at best arbitrary. Additionally, these measures suffer from the absence of baselines: they are all relative estimates, having no ground truth to compare to. These issues are substantial barriers to the interpretability of biases in the coverage of news, which can have a very real impact on the readers' world views~\cite{DellaVigna2005TheFN}. The concentration of media ownership also contributes to reinforcing these biases, since consolidating coverage mechanically weakens media pluralism.  The lack of accountability in these issues is an obvious threat to broadcasting diversity and could jeopardize media integrity, aggravating the public's lack of confidence in news sources.\footnote{http://news.gallup.com/poll/212852/confidence-newspapers-low-rising.aspx}

In this work, we establish a methodology to identify and characterize bias in the mainstream media landscape by recognizing it as a manifestation of the selection process performed by a news source. This paves the way for its treatment as a preference problem, well suited to approaches inspired by personalization methods.

We first argue that capturing this bias would require comparing distributions of covered events \textit{across} news sources, as a biased selection of stories from a news media cannot be observed by looking at the source alone. We thus intend to measure by how much a specific source's news selection deviates from another by learning a latent representation of this source's preferences from its observed selection of events. We hypothesize that this representation allows the study of relationships between sources, and sheds light on the factors that guide their decisions. Last, we claim that this representation could be used to reduce the \textit{selection bias}, proposing a method to promote diversity and equality in the coverage of events by selecting a representative subset of sources.

The rest of this work is organized as follows. In Section~\ref{sec:related} we describe relevant work from the literature. In Section~\ref{sec:data} we describe our dataset as well as our experimental setup. In Section~\ref{sec:methods} we describe our computational approach. In Section~\ref{sec:results} we detail the performance of our approach. In Section~\ref{sec:source_selection} we extend our method by proposing a coverage balancing method. In Section~\ref{sec:discussion} we interpret the results more in depth and investigate several case studies.

\begin{table}
\centering
\begin{tabular}{lp{0.7\linewidth}}
\toprule
Notation 			& Description                 							\\ \midrule
$R$       			& Interaction matrix $\in \mathbb{R}^{|S| \times |E|}$  \\
$S$       			& News source set                    					\\
$E$       			& Event set                    							\\
$s_j$     			& Source $s_j \in S$ 									\\
$e_k$     			& Event $e_k \in E$ 									\\
$K$       			& Number of latent factors 								\\
$\beta$   			& Diversity parameter 									\\
$N$      			& Number of selected sources							\\
$r_{s_j e_k}$ 		& Entry in $R$ (for source $s_j$ and event $e_k)$		\\
$\hat{x}_{s_j e_k}$ & Predicted preference of source $s_j$ for event $e_k$ 	\\
$D$  				& Evaluation set	\\
$N$  				& Number of selected sources	\\
\bottomrule
\end{tabular}
\caption{Notation \label{notation-table}}
\end{table}

\section{Related Work}
\label{sec:related}

\xhdr{Media Bias:} The presence of bias, as well as its formal definition, have been widely discussed in the literature. Early work in the domain, by Groseclose et al.~\cite{10.2307/25098770}, highlighted the left-right cleavage in the coverage of several of the major media outlets by computing an ideological score for each of them. Their approach relies on the observed number of citations of several policy groups in the news relative to the mentions of the same groups by several Congress members. 

More recently, Lin et al.~\cite{Lin2011MoreVT} compared the coverage bias between mainstream media and social media, focusing on stories about the 111th US Congress (2009-2011). They reported a slant in terms of political leaning and a geographic bias.

Saez-Trumper et al.~\cite{SezTrumper2013SocialMN} ran an analysis, at a large-scale, of the bias in both traditional press and social media. They considered three types of biases, namely: \textit{gatekeeping bias}, that defines how stories are selected or ignored in the news, \textit{coverage bias}, that measures how visible an issue is in the news and \textit{statement bias}, that quantifies how the tone of an article is slanted toward or against a particular entity. Specifically, they proposed new metrics to measure biases and characterized those three types of biases in mainstream media as well as social media. The authors' attempt to model gatekeeping bias is perhaps the closest to ours but differs in the fact that it uses an unsupervised approach.

The effects of variations in the news landscape have been surfaced by DellaVigna et al.~\cite{DellaVigna2005TheFN} in an observational study measuring the effect of the introduction of Fox News on voting patterns. Their findings suggest that the Fox channel had convinced 3 to 8 percent of its viewers to vote Republican. 

One of the consequences of a biased press is the formation of a figurative \textit{echo-chamber}, an analogy to the acoustic echo-chamber in which sounds reverberate. The analogy sketches a press in which reputable sources go unquestioned and opposing views are censored. Moreover, the homogenization of views inside an \textit{echo-chamber} artificially reinforces the perception of a universally accepted view. \textit{Echo-chambers} have been studied in social media by Wallsten et al.~\cite{Wallsten2005PoliticalBA}, Flaxman et al. \cite{Flaxman2015FilterBE} and Bakshy et al.~\cite{Bakshy2015PoliticalSE}. 

\xhdr{GDELT:} The GDELT database has been used to observe media response to specific topics such as climate change~\cite{Olteanu2015ComparingEC}, peace and conflicts~\cite{Keertipati2014MultilevelAO} and protests~\cite{Qiao2015GraphBasedMF}.

Kwak et al.~\cite{Kwak2016TwoTO} conducted an extensive experiment to compare the two major news datasets, GDELT and EventRegistry and analyzed their data distributions. They remarked discrepancies in terms of scale, as well as in the included news sources, but observed that the two datasets were following a similar distribution in terms of news geography.

\xhdr{Learning:} Matrix Factorization (MF) methods have gained considerable attention in the last decade, especially in the field of recommender systems, possibly accelerated by the \textit{Netflix Prize} to which Koren et al.~\cite{piotte2009pragmatic} proposed an MF-based solution that was later formalized~\cite{Koren2009MatrixFT}. Despite being principally used in online shopping scenarios, MF methods have been adapted to specific problems, for example in the context of music recommendations~\cite{Nanopoulos2010MusicBoxPM}, bartering platforms~\cite{Rappaz2017BarteringBT} or location-based social networks~\cite{Lian2014GeoMFJG}.

Later advances have studied the problem of learning preferences from implicit feedback~\cite{Hu2008CollaborativeFF}, which are signals of interactions such as click-through rate of purchases. Pan et al.~\cite{Pan2008OneClassCF} considered the extreme case of One-Class Collaborative Filtering (OCCF), in which only positive interactions are observed. Rendle et al.~\cite{Rendle2009BPRBP} proposed Bayesian Personalized Ranking (BPR), a pairwise learning method that handles one-class interaction data while directly optimizing a ranking criterion. We will discuss, in Section~\ref{sec:model}, how the coverage of an event could be treated as a one-class learning problem.

Maximal Marginal Relevance (MMR)~\cite{Carbonell1998TheUO} is an information retrieval technique that retrieves documents based on relevance, while enforcing diversity. It balances the two aspects through the use of a tunable parameter. We refer the reader to Section~\ref{sec:source_selection} for a more detailed description.

\xhdr{Research Questions}
Given the work above, several research questions have remained unanswered:
\begin{description}
\item \textbf{RQ1:} How to capture selection bias in news coverage using supervised learning methods?
\item \textbf{RQ2:} Is the learned representation interpretable?
\item  \textbf{RQ3:} How to exploit the learned bias representation to select a set of news sources exhibiting a balanced coverage?

\end{description}


\section{Data}
\label{sec:data}

\begin{figure}
	\begin{minipage}[t]{0.5\linewidth}
    \includegraphics[width=\linewidth]{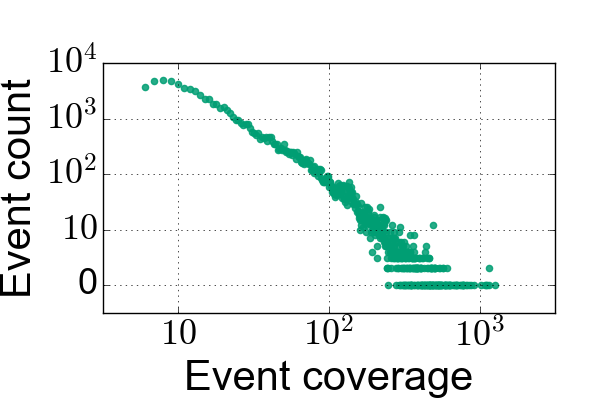}
	\end{minipage}%
	\hfill%
	\begin{minipage}[t]{0.5\linewidth}
    \includegraphics[width=\linewidth]{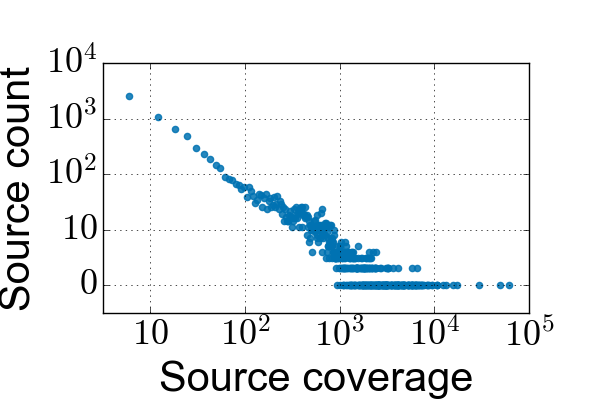}
	\end{minipage}
\caption{Typical distribution of events and sources for a single week.}
\label{fig:distrib}
\end{figure}

In the following section, we describe our data collection process and provide statistics about the resulting dataset. 

\begin{table}[H]
\centering
\begin{tabular}{@{}llllll@{}}
\toprule
 		& Date     			 	 & Sources   	& Events    \\ \hline
Week 1  & 01 - 08 Oct. 	 		 & 9'501 		& 76'966 	\\
Week 2  & 15 - 23 Oct. 	 		 & 9'363		& 88'755 	\\
Week 3  & 25 Oct. - 02 Nov. 	 & 9'741 		& 88'082 	\\
Week 4  & 05 - 13 Nov. 	 		 & 9'714 		& 89'367 	\\
Week 5  & 15 - 23 Oct. 	 		 & 9'961 		& 87'574 	\\ \hline
\end{tabular}
\vspace{0.2cm}
\caption{Meta-data about the 5 selected weeks used in the study.}
\label{tab:weeks}
\end{table}

\subsection{Raw data source}
Recent initiatives, such as the \textit{Global Database of Events, Language, and Tone} (GDELT\footnote{https://www.gdeltproject.org/}) and EventRegistry\footnote{http://eventregistry.org/}, aim to collect, store and process news from all around the world. They have attracted increasing academic attention due to their scale and their temporal coverage. Those initiatives thus represent a unique opportunity to study the specificities of the selection process on a large set of news sources.

Specifically, GDELT is a publicly available catalog of worldwide activities. It actively monitors a wide range of news sources (broadcast, print, and web), recording and annotating global events and their coverage. In this study we extract the necessary data from the GDELT 2.0 Event Database\footnote{https://blog.gdeltproject.org/gdelt-2-0-our-global-world-in-realtime/}: the events table and the mentions table provided by GDELT v2. GDELT events are annotated through standard event coding frameworks~\cite{Leetaru13gdelt:global}, which allow the classification of interactions between world actors. 

The events table references the coverage of events (sampled every 15 minutes) by the sources denoted in the mentions table. The set is then annotated with a best-effort meta-data completion, trying to assign actors, geographical codes, even sentiment scores and categorizing the type of event. Importantly, it assigns a globally unique identifier to each event, which allows continuous tracking of this events' coverage across time and sources.

\subsection{Data processing}
The learning part of our analysis only required to build the interaction matrix between sources and events: we scrape the events and mentions tables to recover which events were covered by which sources in a given timespan. We filter low-count events and sources (sources that have covered less than 5 events, and conversely events covered by less than 5 sources) to limit the impact of the cold-start problem. Fig.~\ref{fig:distrib} and table~\ref{tab:weeks} are computed from our dataset after this preprocessing step.

\section{Methods}
\label{sec:methods}
In the following section, we describe our approach to capture the news selection process. We first describe how to efficiently capture news source preferences in a supervised fashion. We then describe the details of our optimization procedure. Last, we describe the details of our experimental setting as well as our evaluation methodology.

\subsection{Model}
\label{sec:model}
Our method of choice needs to model the decision process of any news source when selecting a subset of events to be covered out of the entire set of available events. We first assume that any given source has a latent preference structure that, in a broad sense, represents its interest in a particular event. If observed, this preference structure would allow ranking any pair of events based on the source interests. Thus, any events covered by a source would be ranked strictly higher than the remaining set of events that it left out. Enforcing this pairwise preference structure in the model offers an elegant way to handle the one-class nature of the data. Indeed, in this scenario, only positive interactions (a source covering an event) are observed. The rest of the interactions is a mixture of real negatives, in the case of the source purposely not covering the event, and missing values, in the case of the source having no information about the event. Thus, the model should be able to handle those unobserved interactions without making strong assumptions about their nature.

Following Rendle et al.~\cite{Rendle2009BPRBP}, we model this decision as a pairwise ranking problem. We train a model to maximize the probability of ranking a positive interaction higher than a negative one. More specifically, we train a model to maximize the following probability for any given news source $s_i$

\begin{equation}
\Pr(e_j>_{s_i}e_k|\Theta),
\end{equation}

where $e_j$ is an event that has been covered by source $s_i$, $e_k$ is an event that has not been covered by $s_i$ and $\Theta$ represents the parameters of an arbitrary predictor.

A predictor that would perfectly model the latent preference structure $>_{s_i}$ of source $s_i$ would thus predict a probability of 1 for $\Pr(e_i>_{s_i}e_j| \Theta)$ and a probability of 0 for $\Pr(e_i<_{s_i}e_j| \Theta)$. Defining $\hat{x}_{s_i,e_j}$ as the predicted score for source $s_i$ and event $e_j$, this can be modeled as $H(\hat{x}_{s_i,e_j,e_k})$ where $\hat{x}_{s_i,e_j,e_k} := \hat{x}_{s_i,e_j} - \hat{x}_{s_i,e_k}$ and $H(\cdot)$ is the Heaviside step function. Note that, in practice, $H(\cdot)$ is not differentiable, and, consequently, is difficult to use with gradient descent methods but can be approximated by a logistic sigmoid function $\sigma(\cdot)$

\begin{equation}
\Pr(e_i>_{s_i}e_j|\Theta) := \sigma(\hat{x}_{s_i,e_j,e_k}(\Theta)) =  \sigma(\hat{x}_{s_i,e_j} - \hat{x}_{s_i,e_k}),
\end{equation}

We described, so far, our modeling of the preference scheme of the observed news channels while delegating the inference to an arbitrary predictor capable of modeling the relationships between sources and events. A suitable predictor should be capable of predicting a score $\hat{x}_{s_i e_j} \in [0,1]$ for every source-event combination where a score of $1$ would represent a high likelihood for a source to cover an event. Modeling this relationship between two sets of discrete components requires a method capable of learning, for every source and every event, a low-dimensional representation that acts as a high-level descriptor of their observed interactions. 

Our insight is that the visible bias is a manifestation of the selection process done by a news source. In other words, the coverage itself can be modeled by a selection process, influenced by a set of real-world factors. In order to model this, we draw an analogy with methods inspired by the field of personalization. Those methods generally rely on the underlying assumption that future interactions of an individual can be predicted by observing users that share a similar behavior, thus needing to establish a relation of the distance between individuals. Following our analogy, we model the news sources as a set of individuals interacting with real-life events.

We select Matrix Factorization (MF) as our method of choice, as it is suited to capture the aforementioned relationship and has produced state-of-the-art results in many personalization applications. We define $R$ as our target matrix of size $\mathbb{R}^{|S| \times |E|}$. The method projects every source and every event in a common low-dimensional space in order to approximate $R$ by learning two low-rank matrices $P$ and $Q$ of size $\mathbb{R}^{K \times |S|}$ and $\mathbb{R}^{K \times |E|}$ respectively, with $K$ being the number of latent factors of the model. As discussed above, the model is learned with a single objective: ranking an observed interaction higher that an unobserved one. A score for source $s_i$ and event $e_j$ can be computed as the dot product of their respective latent-space representations

\begin{equation}
\hat{x}_{s_i, e_j} = p_{s_i}^T \cdot q_{e_j},
\end{equation} 

where $\hat{x}_{s_i, e_j}$ is the predicted score for the given source $s_i$ and event $e_j$ combination and $p_{s_i}$, $q_{e_j}$ are the latent-space representations of source $s_i$ and event $e_j$, respectively.

\subsection{Optimization}

We aim to directly optimize the ranking structure of the problem rather than to provide an accurate reconstruction of the interaction matrix $R$. The BPR optimization scheme introduced by Rendle et al.~\cite{Rendle2009BPRBP} is particularly suited for this type of problem and could be applied to our problem using the following update step

\begin{equation}
	\label{eq:bpr}
	\theta \leftarrow \theta + \alpha \cdot (\sigma(-\hat{x}_{s_i,e_j,e_k}) \
    \frac{\partial\hat{x}_{s_i,e_j,e_k}}{\partial \theta} + \lambda_{\theta}\Omega'(\theta)),
\end{equation}

where $\hat{x}_{s_i,e_j,e_k} = \hat{x}_{s_i e_j} - \hat{x}_{s_i e_k}$, and $\theta$ represents the set of parameters to be learned. $\Omega(\theta)$ denotes a regularizer. We opted for a $\ell_2$ regularization $\Omega(\theta) = \| \Theta \|_2^2$.

\subsection{Experimental Setting}
In order to abstract away temporal dynamics, we proceed to temporally split our data. We select five weeks of interest across 2 months (October and November 2016) in the dataset, which are described in Table~\ref{tab:weeks}. We select one-week chunks to get enough data, and replicate the experiment across the five weeks to measure temporal consistency.

As sources typically cover a highly variable number of events, we adopt a \textit{leave-one-out} methodology to assess the accuracy of the model, with every source having the same weight in the evaluation. Specifically, we constitute our test set by sampling for each source, at random, one event that it covered during the last day of the week.

\xhdr{Reproducibility:} We ran our experiment on a single computer, running a 2.3 GHz Intel Core i7 CPU, using Matlab R2014b. We trained our model with the following parameters: $\alpha = 0.1$, $\lambda_{\theta} = 0.01$, $K = 20$. We found that these were optimal parameters for the proposed problem: the same parameters were used on all 5 weeks. We note that the number of latent factors $K$ did not show significant information gain after $K=20$ dimensions. 

All code will be made available at publication time\footnote{https://selection-bias-www2018.github.io/}.

\subsection{Evaluation} \label{subsec:eval}
Prediction accuracy is not the primary goal of our approach but rather a mean to tune the predictor, in order to avoid under- or over-fitting, and to compare it to various approaches. Since the BPR optimization scheme directly optimizes a pairwise ranking criterion, we select the widely used metric \textit{Area Under the Curve} (AUC)~\cite{shani2011evaluating} as our measure of performance. 

\begin{equation} 
\label{eq1}
\text{AUC} = \frac{1}{\left\vert{D}\right\vert} \sum_{\!\!\!\!\!\!\!\!\!\!\!\!\mathrlap{(s_i, e_j, e_k)\in D}} H( \hat{y}_{s_i e_j} - \hat{y}_{s_i e_k})
		   = \frac{1}{\left\vert{D}\right\vert} \sum_{\!\!\!\!\!\!\!\!\!\!\!\!\mathrlap{(s_i, e_j, e_k)\in D}} H( \hat{x}_{s_i e_j e_k}), 
\end{equation}

where $H(\cdot)$ is the Heaviside step function (the latter formula uses the notation introduced in Section~\ref{sec:methods}) and $D$ is our evaluation set composed of one triplet $(s_i, e_j, e_k)$ per source where $s_i$ is a source, $e_j$ is a randomly sampled event that has been covered by source $s_i$ and $e_k$ is a randomly sampled event that source $s_i$ has not covered. This metric assesses the ability of the predictor to correctly rank a positive interaction withheld during training against a random negative example. An ideal predictor would obtain a score of $AUC=1$, while a random selection would output a score around $AUC=0.5$.

We compare our method to two common baselines used in recommendation problems: popularity and nearest-neighbor methods~\cite{Ricci}. Popularity based methods simply rank the events based on the amount of coverage they receive. Nearest-neighbor methods infer a source's coverage from the coverage of its closest peers: the intuition is that congruent sources should exhibit similar coverage of the event space. We chose the k-Nearest Neighbors ($k=10$) method for this baseline, using the Jaccard distance metric.

\section{Results} \label{sec:results}
In this work, we propose a supervised learning method, which presents the advantage of allowing the explicit evaluation of the quality of our model. We propose that the coverage prediction accuracy yields an adequate estimate of the learned embedding's quality. Indeed, reconstructing the interactions should only be possible if the latent factors captured sufficient information about how sources select the events they cover. This type of evaluation is not feasible with an unsupervised method (e.g. PCA, SVD~\cite{shlens2014tutorial}), which requires expert intervention to judge the quality of results and interpret them. 

\begin{figure}[H]
	\includegraphics[width=\linewidth]{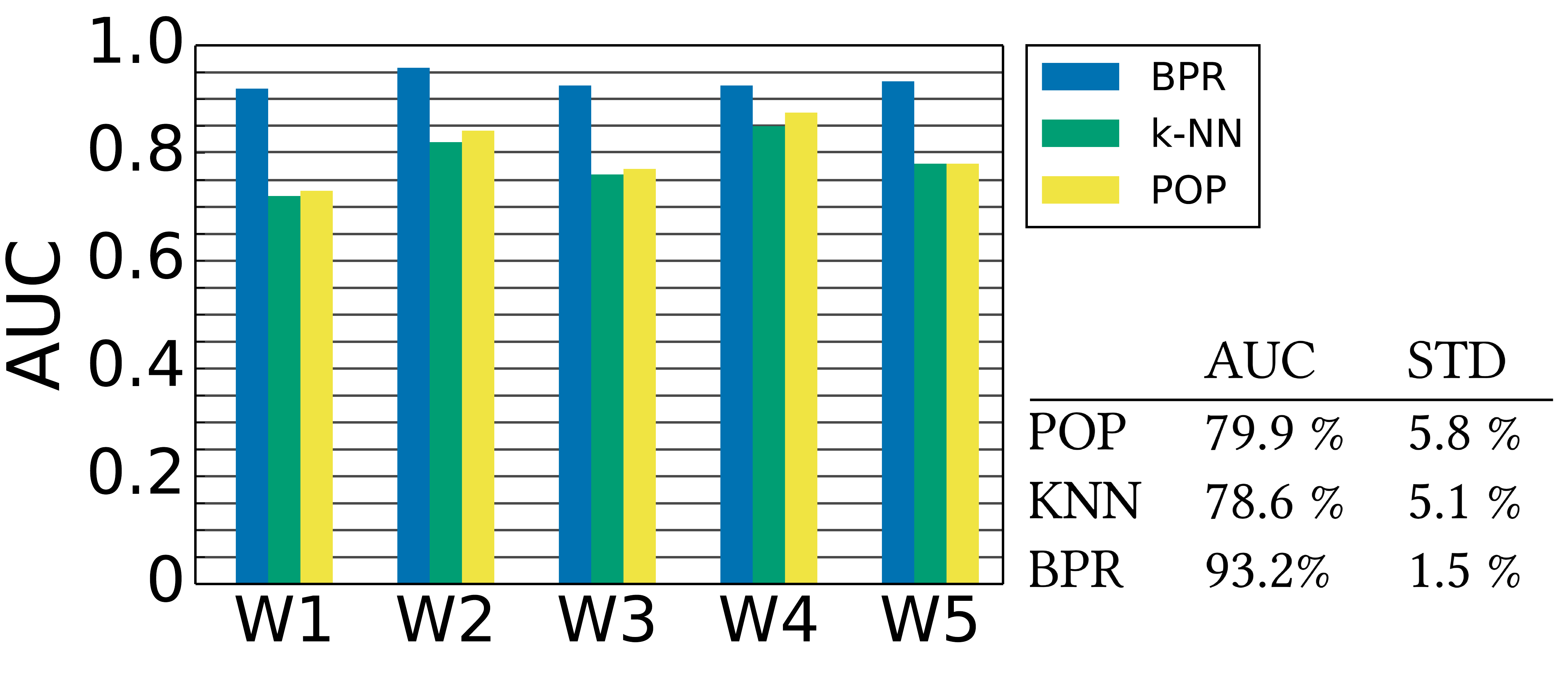}
    \caption{Results with AUC as a performance metric. \textnormal{Results are shown per week. We show the averaged score as well as the standard deviation of the results obtained over the 5 weeks.}}
    \label{fig:auc}
\end{figure}

We reference the results in Fig.~\ref{fig:auc}, which shows higher prediction accuracies compared to the selected baselines.

\begin{figure*}
	\includegraphics[width=\linewidth]{final}
	\caption{Source agglomerations in latent space (best seen in color) \newline \textnormal{\textbf{Left:} After investigation, we observe clusters explainable by the publishing structure of sources in the cluster: all are part of a publishing network, such as the public radio network (\#20: \textit{left, bottom}) or are all owned by a larger commercial entity (\#7: \textit{left, top}, \#18: \textit{left, center}). \textbf{Center:} Position of the sources in latent space, reduced in dimensionality with t-SNE~\cite{Maaten2008VisualizingDU}. An unsupervised cluster learning method (DBSCAN~\cite{Ester1996ADA}) is applied to show agglomerates of sources that are similar in the latent space. 24 clusters are extracted in this example (Week 1). Visual inspection allows interpretation through the discovery of the biases detailed in~\ref{subsec:clusters}. \textbf{Right:} We notice several geographical clusters, three of which are detailed here: a cluster of Indian news sources (\#3: \textit{right, top}), a cluster of Canadian news sources (\#8: \textit{right, center}) and a cluster of sources from Great-Britain and Ireland (\#11: \textit{right, bottom}).}}
\label{fig:clusters}
\end{figure*}

\section{Source Selection} \label{sec:source_selection}
In the following section, we describe how the apriori knowledge produced by our model can be exploited in the context of news selection, that is the problem of selecting $N$ sources from a large and heterogeneous set. In this scenario, the selection of news sources should be done such that the resulting subset exhibits two desirable properties that makes it representative of the worlds' daily events distribution. First, the news sources should be picked in order to foster diversity. Intuitively, the resulting set should cover a large spectrum of the news while minimizing concentration around a small set of events, thus reducing the effect of the so-called \textit{echo-chamber}~\cite{Wallsten2005PoliticalBA}. Second, the resulting set of news covered by the selected sources should retain a large proportion of the most actively covered events, ensuring comprehensive coverage of the event space.

Without an accurate way of modeling the inter-relationship between sources, picking a representative subset of media can be difficult. Indeed, the main criterion of selection would have to come from side-information, e.g. the reputability of the source or its level of activity, etc. Therefore, we propose to exploit the knowledge gained from our model to guide this selection.


We adapt to our scenario a standard diversity-promoting retrieval method, Maximum Marginal Relevance (MMR)~\cite{Carbonell1998TheUO}. MMR is an iterative procedure that establishes a ranking of elements based on two criteria: a relevance score, that is application-specific and has to be defined, and a diversity measure of the retrieved set of elements. MMR balances the two aspects with a tunable parameter $\beta$. At each step, MMR selects the source to be added to the results set based on the relevance of the source, that we define as being comprised in the interval $[0,1]$. This score is then weighted to include results with minimal similarity to the current retrieved set, thus ensuring its diversity. The procedure ranks the sources iteratively based on the following score function

\begin{equation}
\label{eq:MMR}
	\text{MMR}(s_i) :=\beta*\text{relevance}(s_i) - (1-\beta)*\underset{s_j \in B}{\operatorname{max}}\big[ \text{sim}(s_i,s_j)\big],
\end{equation}

where $\beta$ is a parameter that controls the strength of the diversification and $B$ is the set of elements already selected (the first pick is thus based on relevance only). With a $\beta$ value of 1, the ranking is based on relevance only, while with a $\beta$ value of 0 the ranking is the most diverse set of items possible achievable in a greedy fashion. The formulation of equation~\ref{eq:MMR} requires a measure of similarity between sources. After experimenting with different options, we obtained satisfactory results by using $\text{sim}(s_i,s_j)=1/dist(p_i,p_j)$ as our measure of similarity, with $dist(\cdot)$ being the Euclidean distance between $p_i$ and $p_j$, two sources' latent representation vectors. We use as a relevance function the activity level of the source, i.e. the number of articles published by the source (see Section~\ref{sec:applic_news_sel}).

\begin{figure}[H]
\label{fig:illustr_MMR}
\begin{minipage}[t]{0.31\linewidth}
    {%
    \setlength{\fboxsep}{0pt}%
    \setlength{\fboxrule}{0.1pt}%
    \fbox{\includegraphics[width=\linewidth]{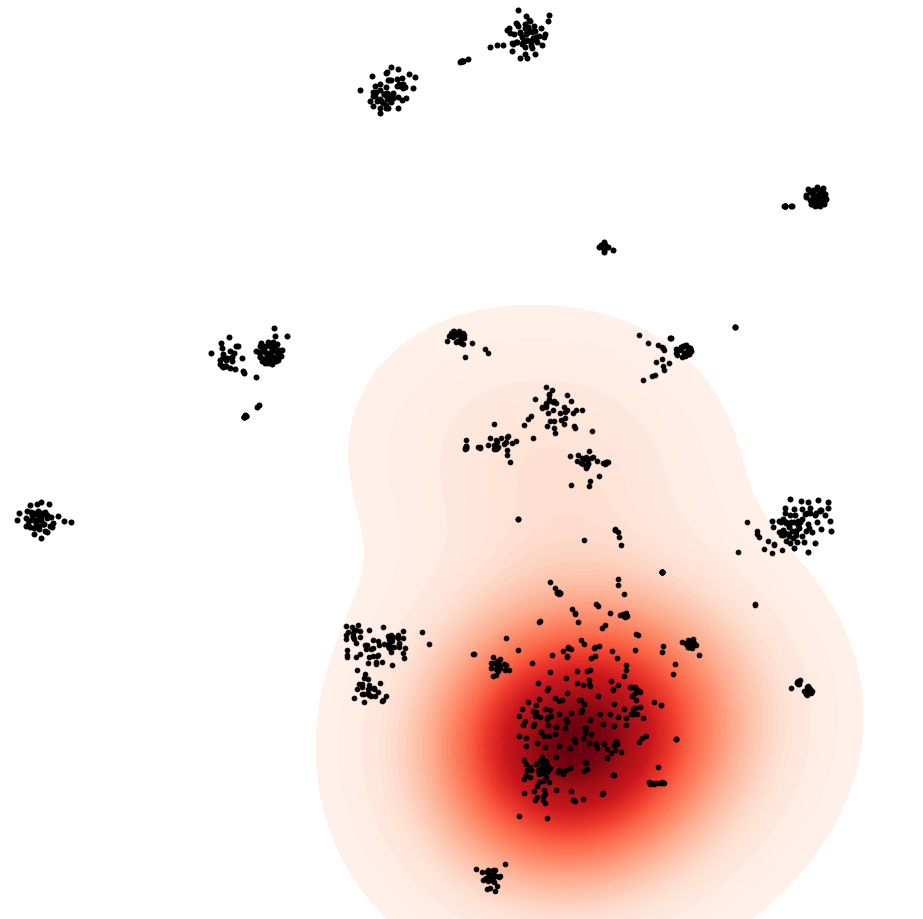}}%
    }%
    \caption*{\textnormal{original}}
\end{minipage}%
\hfill%
\begin{minipage}[t]{0.31\linewidth}
    {%
    \setlength{\fboxsep}{0pt}%
    \setlength{\fboxrule}{0.1pt}%
    \fbox{\includegraphics[width=\linewidth]{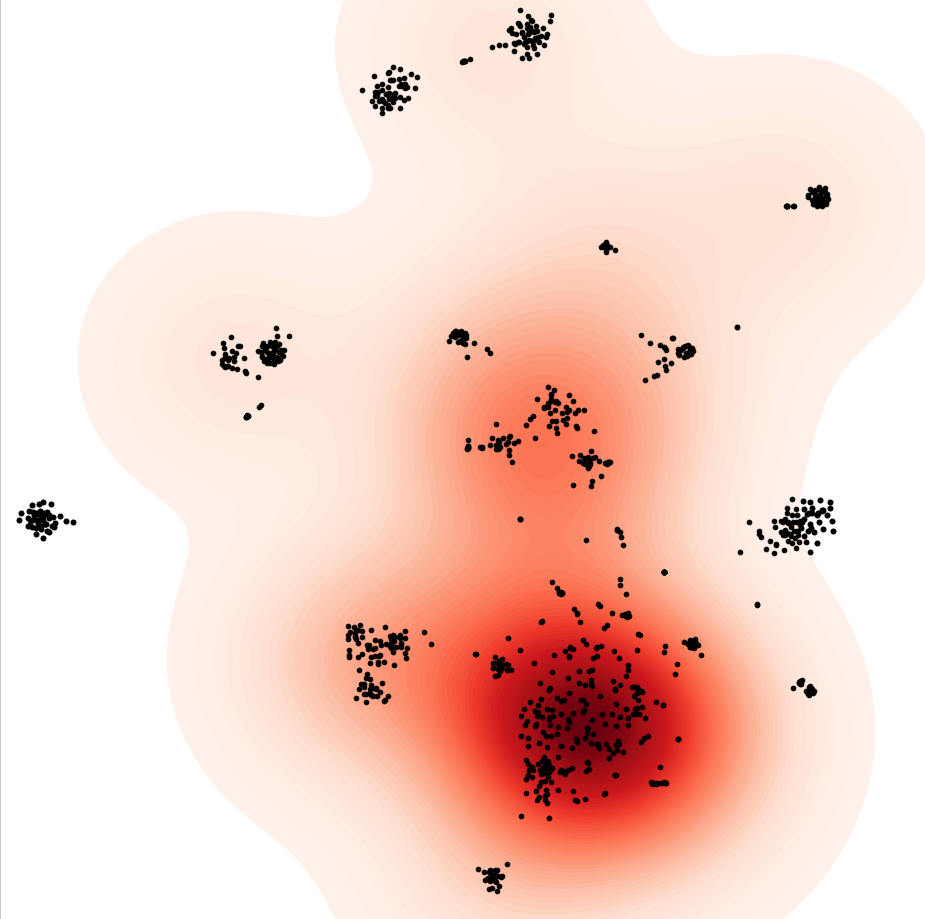}}%
    }%
    \caption*{$\beta = 0.75$}
\end{minipage}
\hfill%
\begin{minipage}[t]{0.31\linewidth}
    {%
    \setlength{\fboxsep}{0pt}%
    \setlength{\fboxrule}{0.1pt}%
    \fbox{\includegraphics[width=\linewidth]{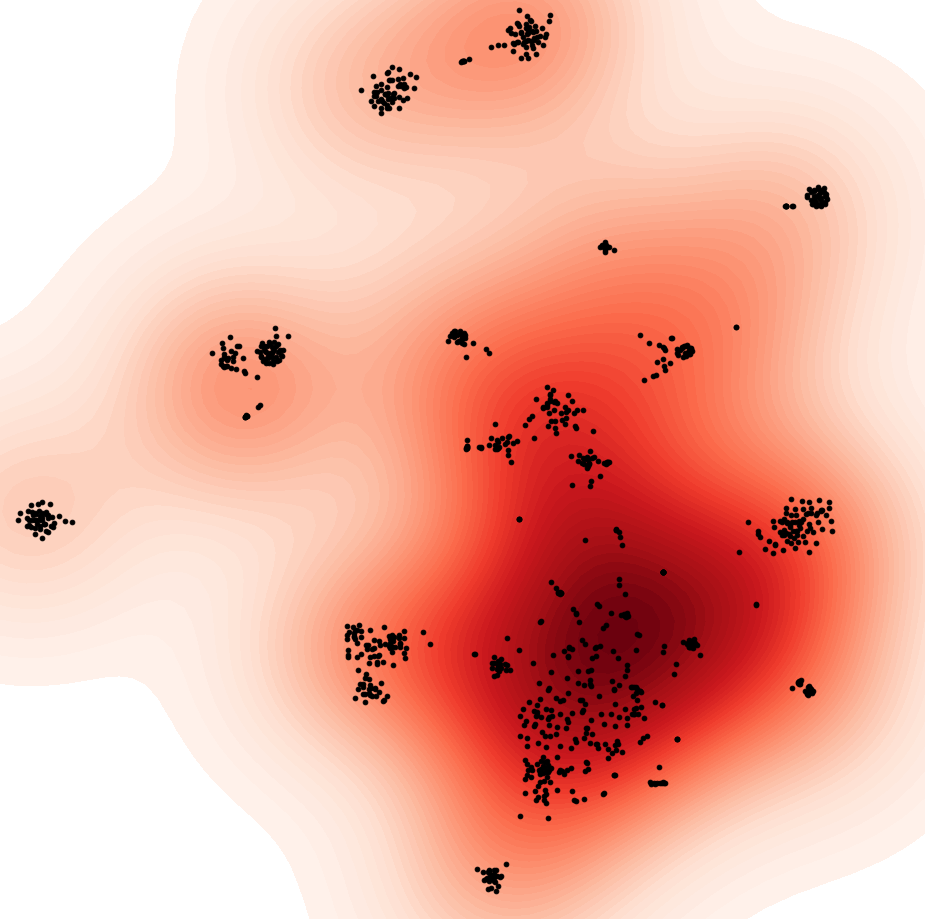}}%
    }%
    \caption*{$\beta = 0.5$}
\end{minipage}
\caption{\textnormal{We illustrate the effect of the $\beta$ parameter on week 5 of our dataset on a query of size N=100
. Sources' positions in latent space are displayed as single dots. We overlay the density (gaussian KDE) around the sources contained in the selected subset. The original selection picks sources solely based on their level of activity ($\beta$=1). The center and right figures have nonzero values of $\beta$ which diversifies the selection of sources.}}
\end{figure}

\section{Discussion} \label{sec:discussion}
In the following section, we consider the results of our experiments. We first discuss the method's predictive performance. Then, we analyze the resulting representations yielded by our approach, providing ways of explaining the observed variance. Last, we describe the results of leveraging this representation with our method to promote diversity in a news source selection problem. 

\subsection{Coverage prediction accuracy}
As mentioned in Section~\ref{sec:results}, our method of choice presents the advantage of supervised learning procedures, in that it provides a measure of the accuracy of the predicted coverage. Therefore, it allows the comparison with other types of personalization techniques. We select two baselines: the raw popularity of the events and k-Nearest Neighbor (k-NN). Popularity based methods are not personalized: they simply rank the events based on the amount of coverage they received. We show that we can outperform this method as a result of the personalization of the coverage prediction. We also compare to a personalized method, k-NN, and observe that our method achieves better accuracy, due to the fact that it is also parameterized. We report a score (AUC) greater than $90\%$ for the 5 selected weeks. We also observe less variability across the weeks in the results obtained from our method of choice.

\subsection{Leveraging representations to uncover biases} \label{subsec:clusters}
The methodology described in Section~\ref{sec:methods} yields latent-space representations of the source preferences, i.e. a low-dimensional description of the selection bias. By investigating the distances between sources in this preference space, we uncover interesting correlations between them, indicating the presence of a common bias. We also apply standard unsupervised clustering methods to explicitly group sources together. While the measures are done in the latent space, we project these vectors down to 2 dimensions for visual inspection. 

Since the structure arises directly from the coverage we can extract factors of the bias, such as those that we mentioned in Section~\ref{sec:intro} (geographic relationships, thematic regards, higher-order structures, ...) despite them not always being evident to the inexperienced eye (for example broadcast affiliates owned by larger structures which are not reflected in branding).

\xhdr{Geographic proximity:} 
The simplest similarity between sources comes from their geographic proximity: local or national sources orient their coverage to their respective scales. Hence sources with similar geographic dependencies should present similarities in their coverage, and be close together in the latent space. This effect is indeed captured by our method, as shown in Fig.~\ref{fig:clusters}, \textit{right}. This geographic relationship between sources is confirmed by the proximity of regional sources, such as \texttt{prokerala.com} and \texttt{newkerala.com}, two sources from the region of Kerala in India: they are in a cluster of Indian news sources, but are also close together in the latent space as they cover national and regional news. The same effect is visible in a portion of cluster \#8, with sources from British Columbia, Canada, being close together (\texttt{westerleynews.com} and \texttt{bclocalnews.com} are shown here).

\begin{figure*}
\begin{minipage}[c]{0.3\linewidth}
    \includegraphics[width=\linewidth]{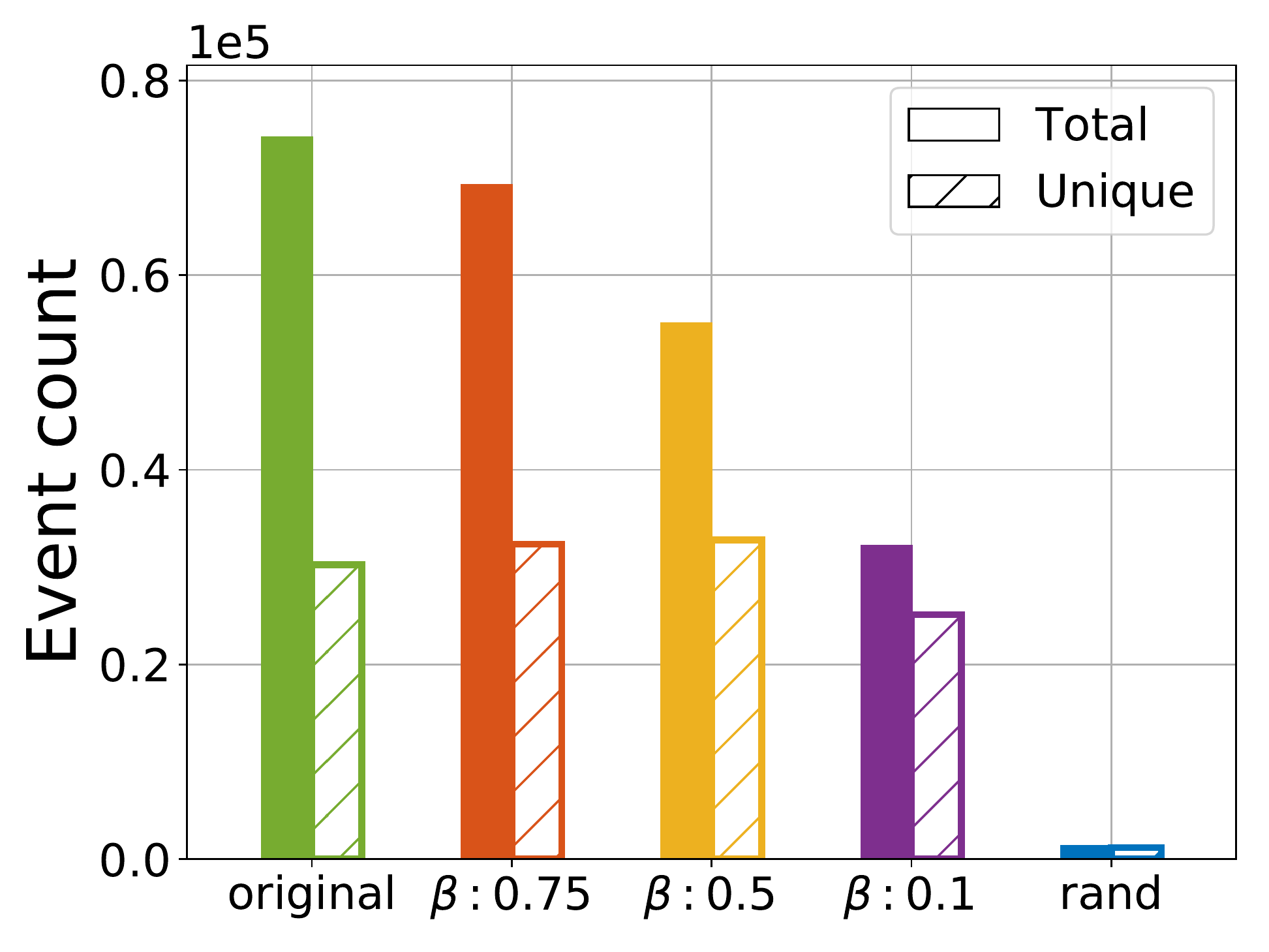}
\end{minipage}%
\hfill%
\begin{minipage}[c]{0.32\linewidth}
    \includegraphics[width=\linewidth]{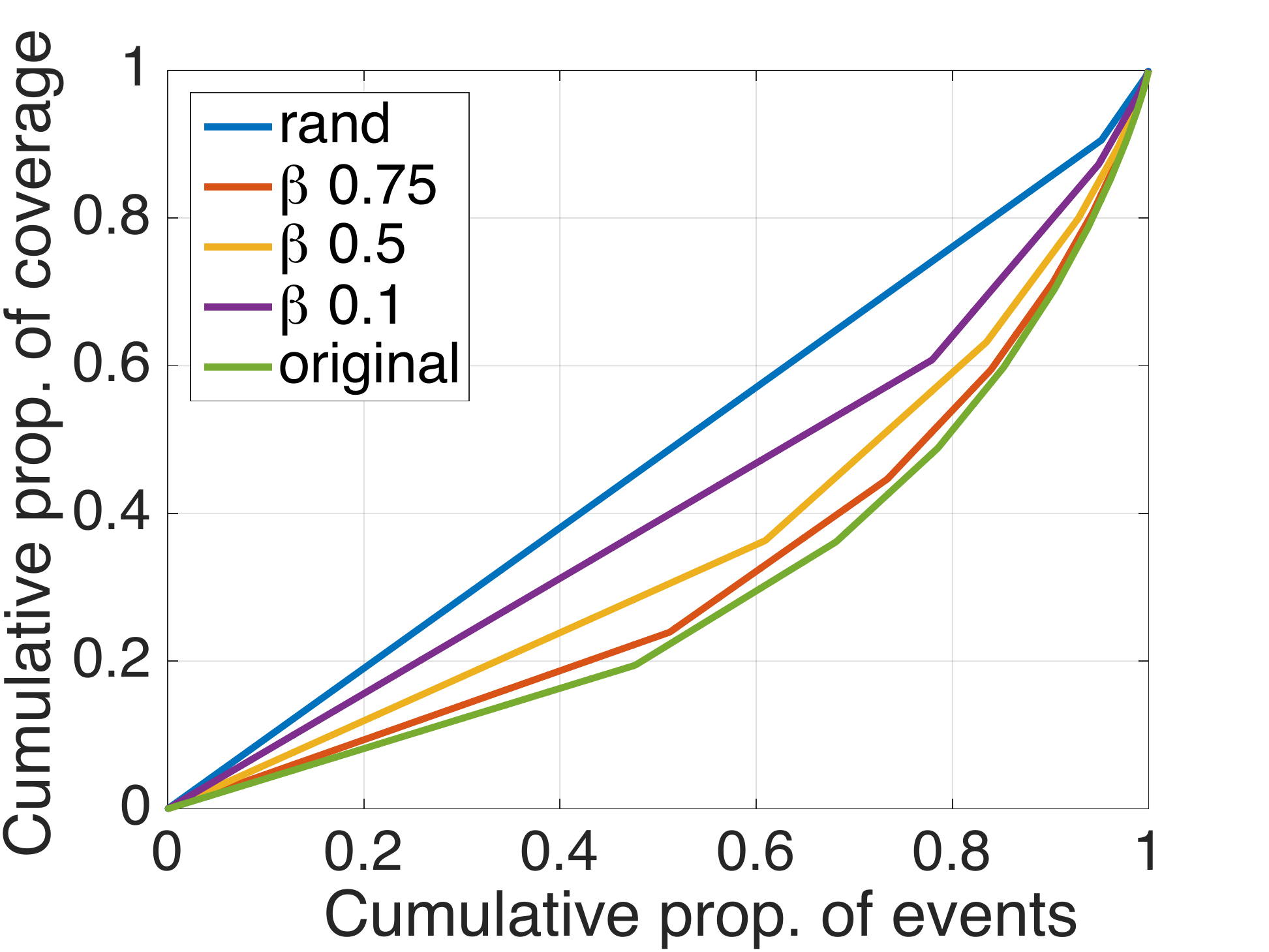}
\end{minipage}
\hfill%
\begin{minipage}[c]{0.32\linewidth}
    \includegraphics[width=\linewidth]{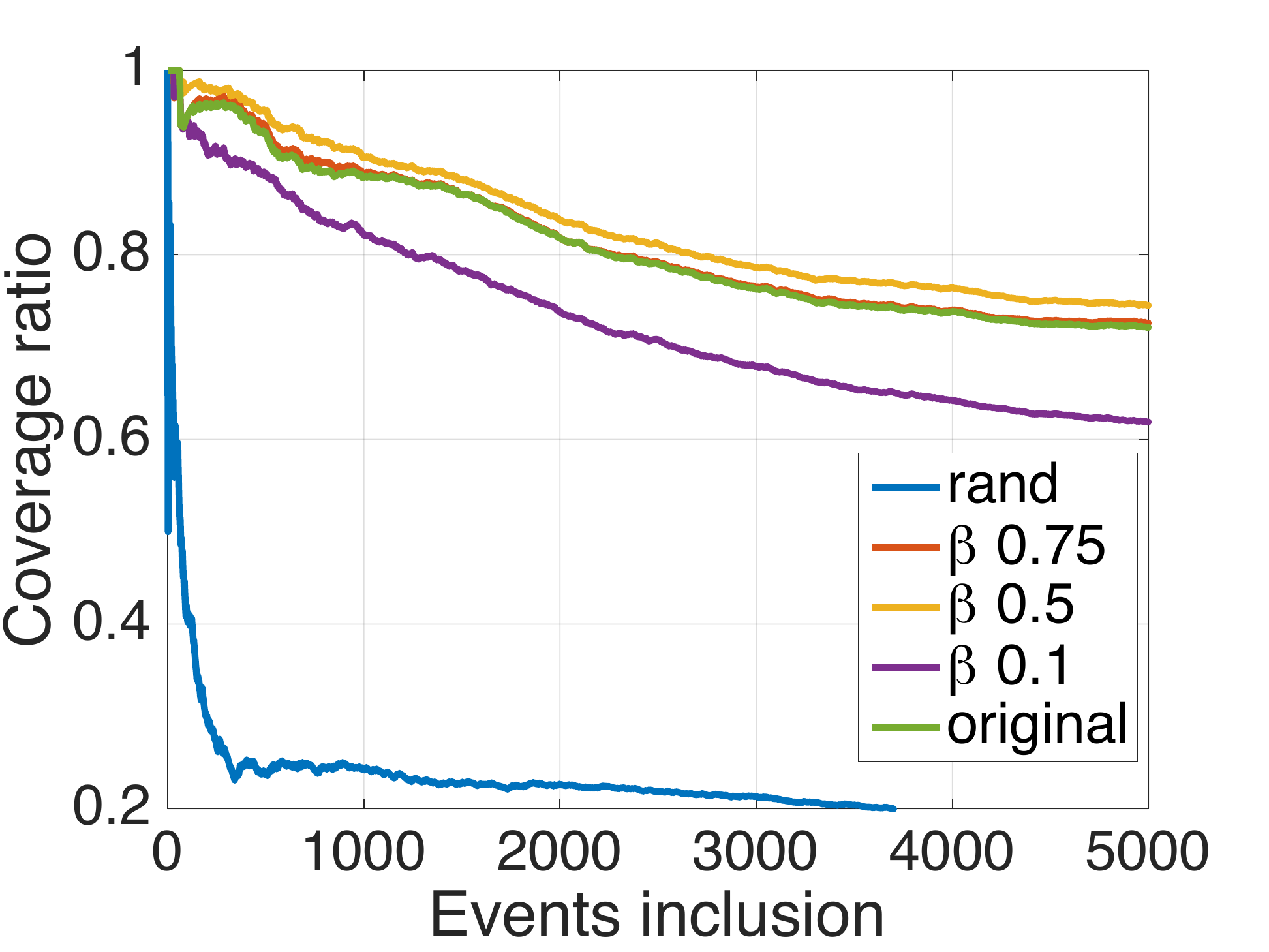}
\end{minipage}
\vfill%
\begin{minipage}[c]{0.3\linewidth}
    \includegraphics[width=\linewidth]{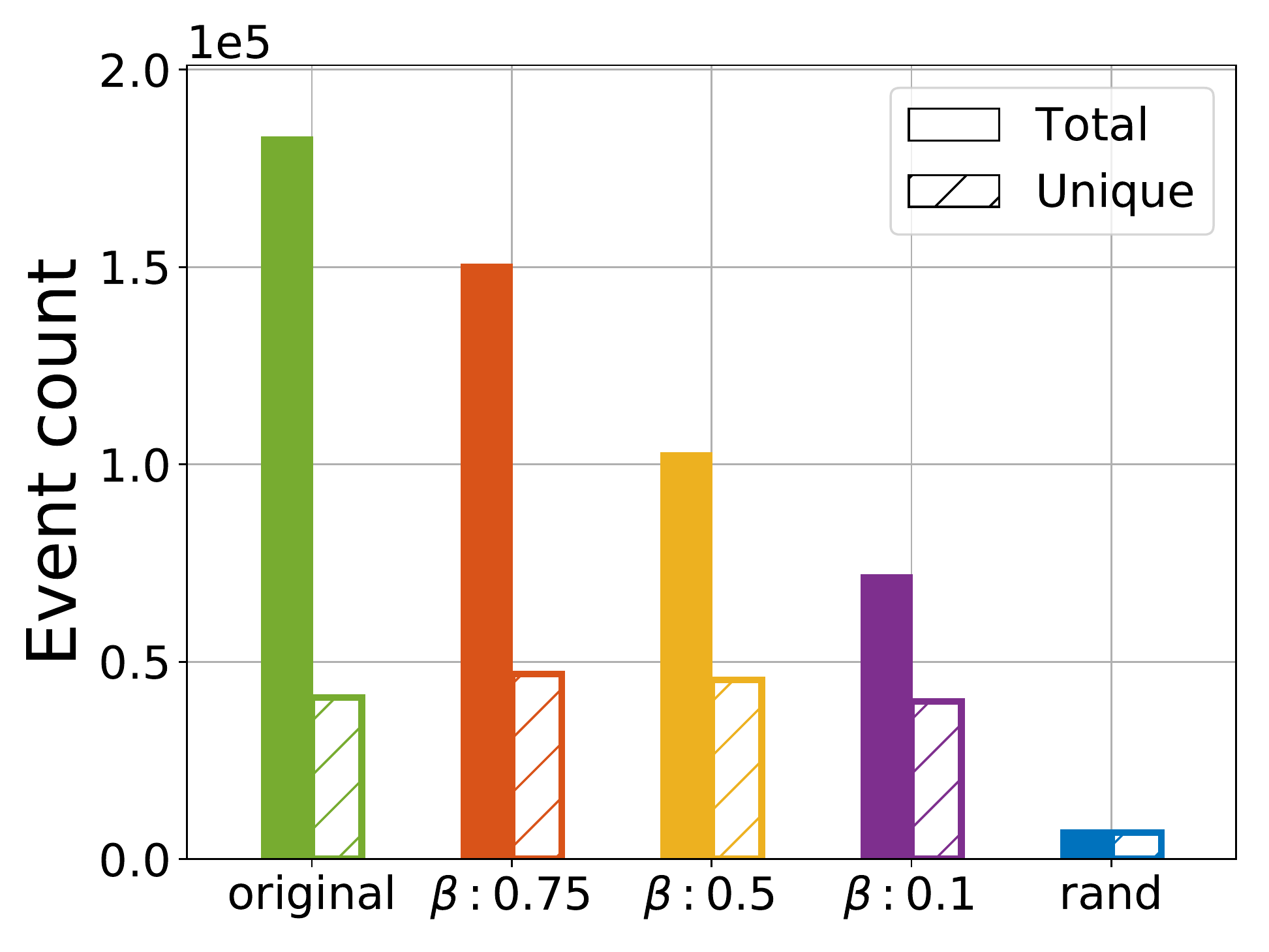}
\end{minipage}%
\hfill%
\begin{minipage}[c]{0.32\linewidth}
    \includegraphics[width=\linewidth]{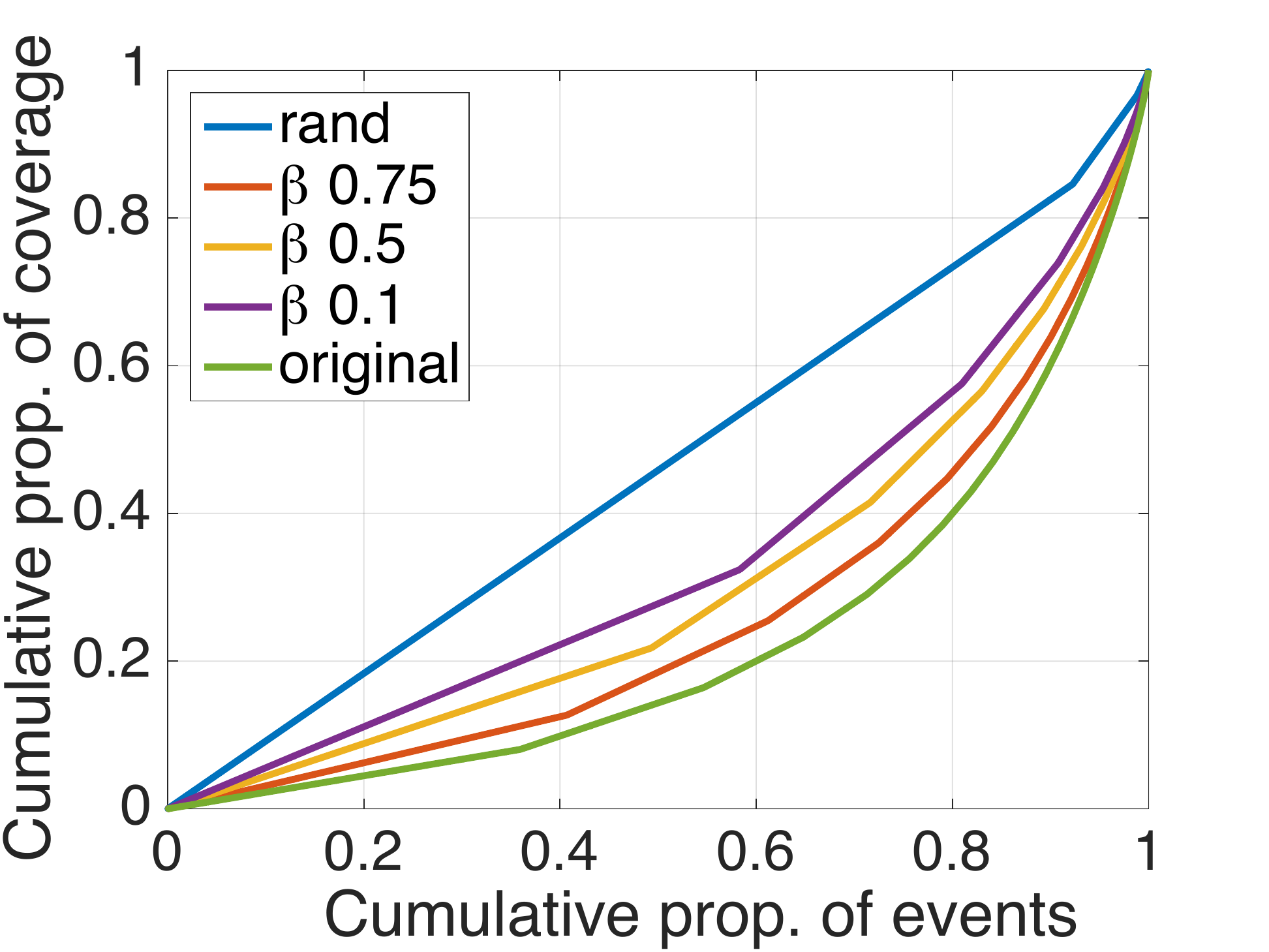}
\end{minipage}
\hfill%
\begin{minipage}[c]{0.32\linewidth}
    \includegraphics[width=\linewidth]{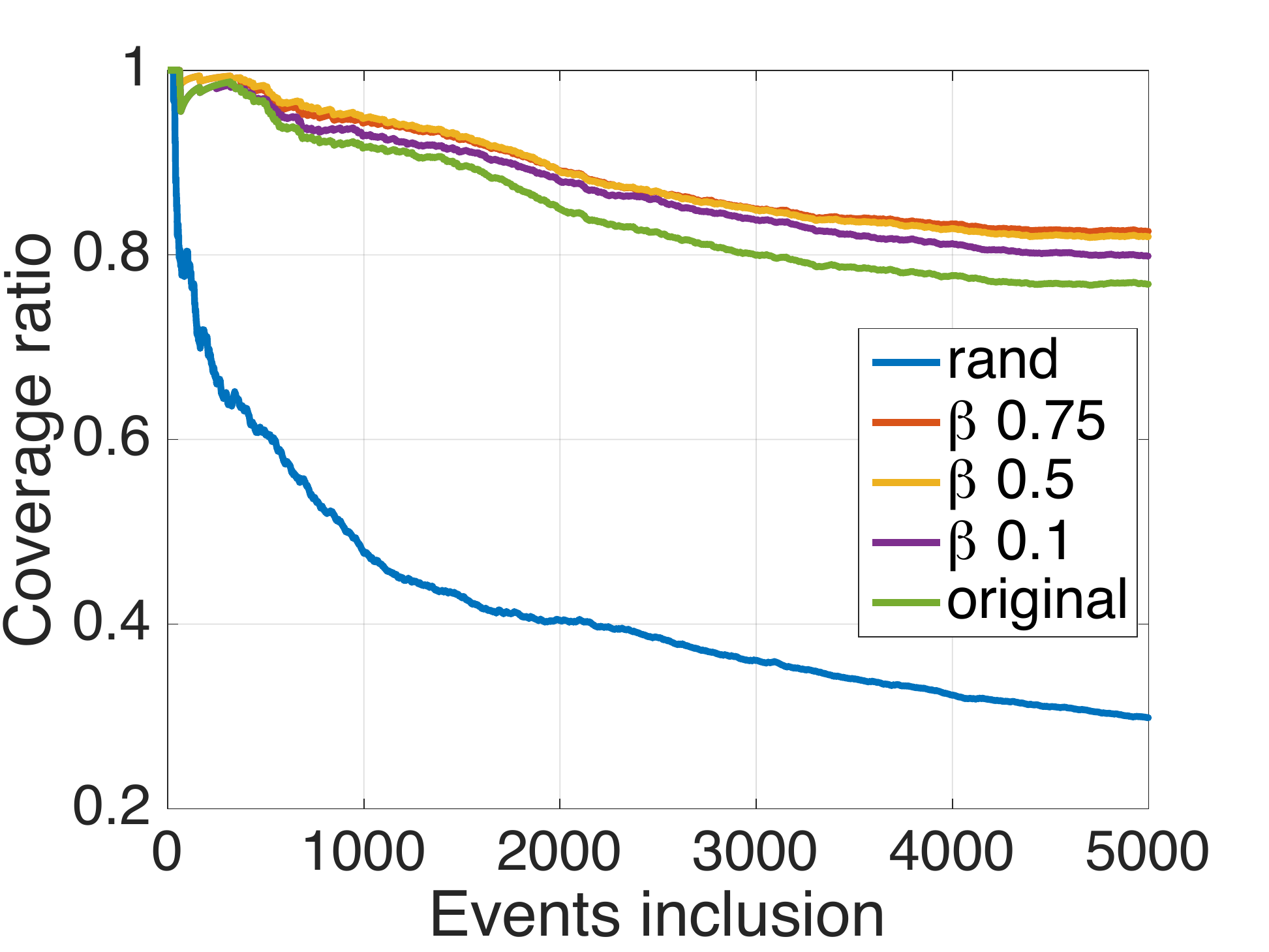}
\end{minipage}

\caption{top-25 sources selection (first row) and top-100 sources selection (second row). \textnormal{We report the coverage produced by the original ranking (ranked by the number of articles published), the same ranking with the diversity constraint for different values of $\beta$, and the coverage of a random selection of sources.} Left: \textnormal{Number of articles covered (total and unique) by the selected subset of news sources.}
Center: \textnormal{Lorenz curves of the coverage received by individual events in the selected subset of news sources.}
Right: \textnormal{Proportion of the top-5000 most discussed events of the week included in the coverage of the selected subset of news sources. For example, a top-100 on the x-axis represents the percentage of the 100 most covered events in the entire set that have been covered at least once by the selected subset of news sources.}
}
\label{fig:source_select}
\end{figure*}

\xhdr{Affiliation and ownership:} Local news sources are an essential part of the news coverage network, most notably in rural areas where they represent one of the only sources of information with a granularity level fine enough to cover very local events. While it is to their advantage to also provide general news coverage to their readers (national or international news), they usually lack the resources to be involved in the treatment of events at that scale. Hence a common method has long been to agglomerate into larger organizations: groups of local news sources dedicating a fraction of their budgets to pool the coverage between them, forming a broadcast syndication network~\cite{massComLorimer94}. 

Note that these groupings are not necessarily horizontal: they can also be the fruit of consolidations through mergers or acquisitions by larger organizations (the Pew Research Center estimates that the five largest broadcast companies now own 37\% of local television stations in the United States\footnote{http://www.pewresearch.org/fact-tank/2017/05/11/buying-spree-brings-more-local-tv-stations-to-fewer-big-companies/}). In cluster \#18, we show a group of sources all owned by the same corporate structures, formed by a wave of acquisitions in the local news space.  

These larger structures are not always obvious at a glance. Many familiar networks are present in the list of sources shown in Fig.~\ref{fig:clusters}, such as the American Broadcasting Company (ABC\footnote{abcnews.go.com}) (\texttt{abc22now.com}), Columbia Broadcasting System (CBS\footnote{cbs.com}) (\texttt{cbs12.com}) or even Fox\footnote{foxnews.com} (\texttt{okcfox.com}) but none of these are actually operated by the network their name suggests: they are all operated by the same broadcast entity.

\xhdr{Medium:} Some of the larger structures that form are driven by platforms based on similar media. Cluster \#20 brings together a network of public radio stations. They are usually affiliated with one or several organizations such as NPR\footnote{npr.org}, Public Radio International\footnote{pri.org} or American Public Media\footnote{americanpublicmedia.org}, all of which are non-profit entities exchanging content to form a radio syndicate. 

A few observations are left as side-notes. First, we report the clusters discussed in this section to be largely consistent throughout the 5 selected weeks. We report an average Pearson correlation of $\boldsymbol{0.82}$ between the pairwise distances in embedding space of the top-1000 most active sources across the 5 weeks. Second, we did not observe any clear left-right cleavage, and, therefore, do not report on it.

\subsection{Application to source selection}
\label{sec:applic_news_sel}
In this section, we develop the results obtained by the proposed method in the context of source selection. In particular, the properties of their combined coverage of the event space is of our interest. A skewed selection of news sources could induce side-effects. The selected sources could cover a too-small or non-representative portion of the event space by focusing on a few highly discussed topics. As a consequence, those events would be overrepresented while other topics of importance would be drowned. Therefore, we discuss the results of the news selection problem with respect to two aspects. First, we report a metric of coverage equality received by the events. An \textit{egalitarian coverage} should give a similar importance to all events treated by our selected sources. Second, we report the ability of the method to retain the most actively covered events in the set.

We first select a subset of $N$ news sources based on a ranking criterion that does not require any side-information: their respective levels of activity. This naive approach ensures the resulting selection to include the largest possible number of articles. We therefore expect it to contain a wide spectrum of events. We then compare this coverage with the one produced by a ranking with the additional diversity constraint presented in Section~\ref{sec:source_selection}. We report that a skewed attention in the original ranking of sources provides a ratio of $\#\text{events} / \# \text{articles}$ of $\boldsymbol{0.41}$ for top-$25$ and $\boldsymbol{0.22}$ for top-$100$. This ratio suggests a lot of repetitions around the same subset of events. However, we observe this effect being mitigated by the re-ranking procedure. For example, we obtain a ratio of $\boldsymbol{0.60}$ for top-$25$ and $\boldsymbol{0.44}$ for top-$100$, by fixing the value of the $\beta$ parameter to $0.5$. A more detailed view of this discrepancy, and its mitigation, is shown in figure~\ref{fig:source_select} (left). 

This ratio gives an indication of the overall novelty provided by a set of sources. However, it does not show the unequal treatment of the event, which we have hypothesized. If we consider the coverage of news sources as a budget of attention, we observe the attention income that every event receives. In fact, the Lorenz curves (figure~\ref{fig:source_select} center) indeed reveal the attention budget of the press being spent unequally for a selection of the most active sources. We report that this effect is also mitigated by the proposed approach. We also estimate the imbalance in the resulting coverage in statistical terms using the GINI coefficient which measures the inequality of a distribution. A perfectly egalitarian coverage would have a GINI coefficient of 0, meaning all events receive equal attention. For a selection of 25 sources, we obtain a GINI coefficient of $\boldsymbol{0.79}$ that reduces to $\boldsymbol{0.74}$ after re-ranking ($\beta=0.5$). Similarly, for a selection of 100 sources, we obtain a GINI coefficient of $\boldsymbol{0.78}$ that reduces to $\boldsymbol{0.68}$ after re-ranking ($\beta=0.5$).

Although equality in the coverage is a desirable property, we cannot sacrifice the total coverage to achieve an egalitarian distribution: this would mean discarding too many important events for the coverage to be meaningful. Hence we also report the propensity of our selected subset to retain events of importance, as shown in Fig.~\ref{fig:source_select} (\textit{right}). We ranked the event by importance, the top events being the ones that have been covered by a larger number of sources during the week. We show the resulting selection of news sources includes a larger proportion of the most discussed topics despite covering a smaller set of unique events.

The last point of our discussion treats of the balance between coverage equality and top-event retention. The choice of the $\beta$ parameter is a trade-off between the two aspects that could be fixed through numerical analysis or include human judgment. However, a value of $\beta=0.5$ allows to substantially reduce the imbalance, while still including a larger proportion of top events in the resulting coverage.

\section{Conclusions and Future Work}
We studied the presence and nature of selection biases in the context of news coverage. By treating the event selection as a preference problem, motivating the application of methods inspired by personalization systems, we reported distinct and interpretable communities of news sources by learning from their coverage alone. Notably, these agglomerations present high cohesiveness suggesting the media landscape to be hierarchical. We further leveraged the learned representations to propose a methodology producing more diverse and egalitarian coverage of the news. Moreover, we reported this re-ranking procedure to preserve a larger proportion of the most discussed events compared to a simple selection of highly active sources.

\vfill \eject

The learned representations shed light on many real-world relationships between news entities, which in turn influence the coverage of the news by these sources. Notably, we detected geographic dependencies, conserving even regional links, as well as same-medium sources without the use of side-information. We also report the ability to extract non-trivial relationships, such as affiliations, broadcast syndications and even the inclusion of sources in a corporate network. The identification of these non-obvious structures is an important step towards the transparency needed to restore public trust in the reporting process. Furthermore, our diversity-promoting news coverage selection method can hinder the effect of having many sources but few voices, favoring media pluralism, another essential block in the effort towards more trustworthy information sources.

Several points are left for future work. First, our methodology treats sources and events as discrete components with no external information attached, and, thus, could suffer from the cold-start problem. A future research effort could integrate side-information into the optimization procedure to limit its effect and increase performance. Second, we analyzed the problem of selecting sources but abstracted the selection of sources. Indeed, in this work, articles from different sources treating the same event are considered equivalent. A future research scenario could differentiate news articles from different sources, from different views, by capturing semantic information from the articles. Last, we believe that the meta-analysis would greatly benefit from the insight of domain experts to uncover non-evident relationships within the formed clusters.

\clearpage

\bibliographystyle{ACM-Reference-Format}
\bibliography{sample-bibliography} 

\end{document}